\documentclass[conference]{IEEEtran}
\IEEEoverridecommandlockouts
\usepackage{url}
\usepackage{tabularx}
\newcolumntype{L}[1]{>{\raggedright\arraybackslash}p{#1}}
\newcolumntype{C}[1]{>{\centering\arraybackslash}p{#1}}
\newcolumntype{R}[1]{>{\raggedleft\arraybackslash}p{#1}}
\usepackage{booktabs}

\usepackage{cite}
\usepackage{amsmath,amssymb,amsfonts}
\usepackage{algorithmic}
\usepackage{graphicx}
\usepackage{textcomp}
\usepackage{xcolor}
\usepackage{subfigure}
\def\BibTeX{{\rm B\kern-.05em{\sc i\kern-.025em b}\kern-.08em
    T\kern-.1667em\lower.7ex\hbox{E}\kern-.125emX}}
\begin{document}

\title{Blood Vessel Detection using Modified Multiscale MF-FDOG Filters for Diabetic Retinopathy\\
}

\author{\IEEEauthorblockN{Debojyoti Mallick\IEEEauthorrefmark{1},
Kundan Kumar\IEEEauthorrefmark{2}, and Sumanshu Agarwal\IEEEauthorrefmark{3}}
\IEEEauthorblockA{\textit{Department of Electronics and Communication Engineering} \\
\textit{Institute of Technical Education \& Research, S`O'A (Deemed to be University)}, 
Bhubaneswar, India \\
Email: \IEEEauthorrefmark{1}debojyotimallick94@gmail.com, \IEEEauthorrefmark{2}erkundanec@gmail.com, \IEEEauthorrefmark{3}sumanshuagarwal@soa.ac.in}}
\maketitle


\maketitle

\begin{abstract}
Blindness in diabetic patients caused by retinopathy (characterized by an increase in the diameter and new branches of the blood vessels inside the retina) is a grave concern. Many efforts have been made for the early detection of the disease using various image processing techniques on retinal images. However, most of the methods are plagued with the false detection of the blood vessel pixels. Given that, here, we propose a modified matched filter with the first derivative of Gaussian. The method uses the top-hat transform and contrast limited histogram equalization. Further, we segment the modified multiscale matched filter response by using a binary threshold obtained from the first derivative of Gaussian. The method was assessed on a publicly available database (DRIVE database). As anticipated, the proposed method provides a higher accuracy compared to the literature. Moreover, a lesser false detection from the existing matched filters and its variants have been observed.
\end{abstract}


\section{Introduction}
\begin{figure}[h]
    \centering
    \subfigure[]{\includegraphics[width = 4cm]{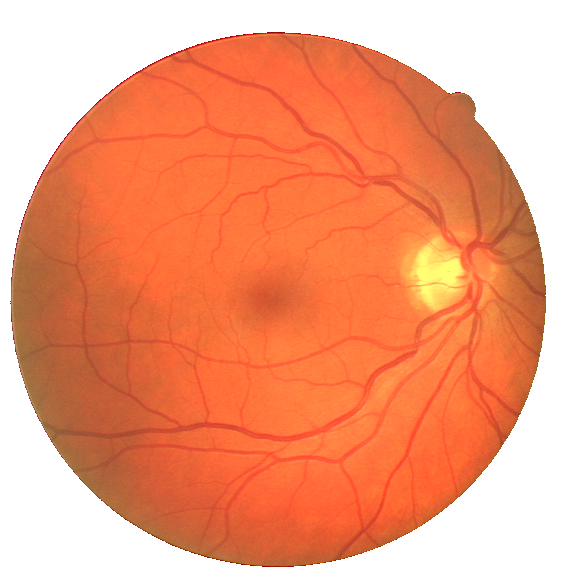}}
    \subfigure[]{\includegraphics[width = 4cm]{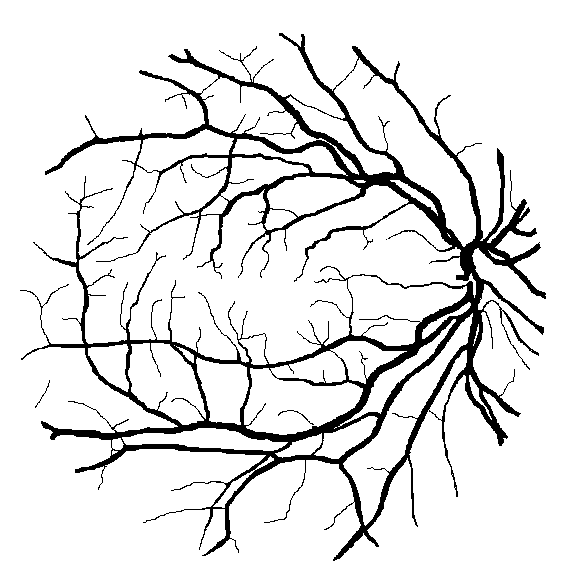}}
    \caption{(a) A representative retinal image (b) with its ground truth.}
    \label{Fig_treeStructure}
\end{figure}

Lately, digital image processing has been a very popular toolbox in the field of medicine, navigation system, agriculture, object tracking, etc.\ \cite{Acharya2005}. Researchers have successfully detected diseases like cancer, diabetic retinopathy, brain tumor, breast calcification, etc.\ using various image processing techniques~\cite{MedicalImaging}. Blood vessel detection (BVD) is an emerging area of research in the field to identify the illnesses like hypertension, arteriosclerosis, diabetes, etc.~\cite{Zhang2009}. Here, the estimates are made by observing a change in the diameter and branching of the blood vessels (a tree-like structure) inside the retina (refer to Fig.~\ref{Fig_treeStructure}). 

The reported BVD approaches can broadly be classified in 4 categories viz., (i) tracking based approach, (ii) morphological based approach, (iii) filtering based approach, and (iv) machine learning approach. Out of these, the first three approaches are based on unsupervised learning; however, the last method is based on supervised learning~\cite{Fraz2012, Almotiri2018}. Nevertheless, filtering based approaches are still very popular~\cite{Kumar2016, Gou2018,Raman2019}. Literature indicates that various proposed algorithms in the domain are based on line detection techniques~\cite{Nguyen2013}, mainly due to similarities in BVD problem and line detection problem. Further, the use of matched filter has been demonstrated~\cite{Chaudhuri1989}. In addition, various thresholding techniques, such as adaptive thresholding, entropy based thresholding, etc., have been applied on matched filter response to obtain the binary map (blood vessel and background)~\cite{Hoover2000}. However, matched filter response fails to distinguish between non-vessels and vessels edges in retinal images~\cite{Zhang2009}. To mitigate the shortcoming, matched filter with first-order derivative of Gaussian (MF-FDOG)~\cite{Zhang2009} and modified matched filter with double-sided thresholding~\cite{Zhang2010} have been proposed. Further reduction in the false positive detection was observed by combining the zero crossing property of the Laplacian of Gaussian with matched filter~\cite{Kumar2016}. 
 However, all of these methods assume that the intensity profile across the blood vessel follow the Gaussian distribution, which may not be true always. On the other hand, matched filter based approach is free from the deceptive constraints.

Given that, here, we propose an automated unsupervised approach based on matched filter for automatic segmentation of blood vessel map from retinal image. 
In this paper, we a) introduce a modified multiscale MF-FDOG filter which process the retinal blood vessels by considering neighborhood pixels of blood vessel having Gaussian profile, b) use top-hat transform followed by contrast limited adaptive histogram equalization (CLAHE) for retinal image enhancement in preprocessing stage, and c) perform the quantitative analysis of the proposed framework on publicly available Digital Retinal Image for Vessels Extraction (DRIVE) database~\cite{Staal2004}. We find that the proposed method significantly improves the true positive detection and reduces the false detection.

In the next section, we discuss the MF-FDOG. The proposed modified MF-FDOG methods is discussed in the section~\ref{ModMFFDOG}. In section~\ref{extSec}, blood vessel extraction framework is presented. We discus the results in section \ref{resultsec}. The conclusion of the work is thereon section~\ref{conclusion}. 

\section{Background}
Matched filter with first-order derivative of Gaussian (MF-FDOG) has been found an effective method to extract the retinal vessels~\cite{Zhang2010}. The MF-FDOG as defined by Zhang \textit{et.al.}~\cite{Zhang2010} is given below.
\begin{equation}
    k(x,y) = \frac{1}{\sqrt{2\pi}\sigma}{\exp\left(-\frac{x^2}{2\sigma^2} \right)-\mu}~~\text{for}~|x|\leq t\cdot \sigma, |y|\leq L/2,
    \label{Eq_MF}
\end{equation}
where $\sigma$ and $\mu$ are the scale and the mean of the matched filter kernel $k$. It is defined as
\begin{equation}
    \mu  = \frac{1}{{2t\sigma }} \cdot \left( {\int\limits_{ - t\sigma }^{t\sigma } {\frac{1}{{\sqrt {2\pi } \sigma }}\exp \left( {\frac{{ - {x^2}}}{{2{\sigma ^2}}}} \right)dx} } \right)
\end{equation}

In equation~(\ref{Eq_MF}), $\mu$ is subtracted from the kernel to obtain the zero mean filter to 
remove the smooth background. Here, $(x,y)$ is the spatial coordinates of the filter which depends on the parameter $t$ and $L$ respectively. $t$ is used to maintain the confidence interval of the Gaussian distribution along the $x$-axis, however, the parameter $L$ is the length of kernel along the $y$-axis. In usual practice, the value of $t$ is set as 3 so that 99\% of the area under the Gaussian curve lies within the range $[-3\sigma,3\sigma]$. Further, the first derivative of Equation~(\ref{Eq_MF}) is
\begin{equation}
    k'(x,y) =  - \frac{x}{{\sqrt {2\pi } {\sigma ^3}}}\exp \left( { - \frac{{{x^2}}}{{2{\sigma ^2}}}} \right)~~\text{for}~\left| x \right| \leqslant t\cdot \sigma,\left| y \right| \leqslant L/2.
    \label{Eq_FDOG}
\end{equation}
Now, we need to estimate the threshold value for the matched filter to implement it effectively. Here, the threshold value for the matched filter is obtained by using response of the image to FDOG. Filter bank for matched filter and FDOG are obtained by rotating the kernels $k$ and $k'$. Filter bank response $H$ and $D$ of retinal image are obtained by apply filter kernel $k$ and $k'$ respectively. The filter response $D$ is averaged using mean filter $W$ of size $w\times w$ as
\begin{equation}
    D_m=D*W
\end{equation}
The normalized response ${\bar D_m}$ is obtained by normalizing each pixel value of the averaged response $D_m$. Then, threshold $T$ is computed as
\begin{equation}
    T=(1+{\bar D_m})\cdot T_c
\end{equation}
where $T_c$ is the reference threshold computed as $T_c=c\cdot \mu_H$, where $\mu_H$ is the mean value of matched filter response $H$, and $c$ is a constant. To obtain the binary vessel map $M_H$, the threshold $T$ is applied to $H$ as
\begin{equation}
    \begin{array}{*{20}{c}}
  {{M_H} = 1}&{H(x,y) \geqslant T(x,y)} \\ 
  {{M_H} = 0}&{H(x,y) < T(x,y)} 
\end{array}
\end{equation}
It can be noticed here that for vessel pixels in retinal image, the threshold value is lower, however, for non-vessel pixels the threshold value is higher. Therefore, blood vessel pixel sustain whereas non-vessel pixels vanished in the binary vessel map due to higher threshold.

\section{Proposed Modified MF-FDOG}
\label{ModMFFDOG}
\begin{figure}[!htbp]
    \centering
    \includegraphics[width=0.07\textwidth]{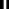}~\includegraphics[width=0.07\textwidth]{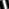}~\includegraphics[width=0.07\textwidth]{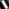}~\includegraphics[width=0.07\textwidth]{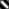}~\includegraphics[width=0.07\textwidth]{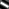}~\includegraphics[width=0.07\textwidth]{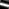}\vspace{4pt}\\\includegraphics[width=0.07\textwidth]{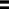}~\includegraphics[width=0.07\textwidth]{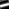}~\includegraphics[width=0.07\textwidth]{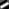}~\includegraphics[width=0.07\textwidth]{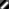}~\includegraphics[width=0.07\textwidth]{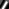}~\includegraphics[width=0.07\textwidth]{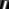}
    \caption{Match filters of size $9\times 9$ at the orientation difference of $15^{\circ}$.}
    \label{fig:matchFilters}
\end{figure}

\begin{figure}[h]
    \centering
    \includegraphics[width=0.07\textwidth]{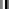}~\includegraphics[width=0.07\textwidth]{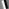}~\includegraphics[width=0.07\textwidth]{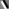}~\includegraphics[width=0.07\textwidth]{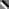}~\includegraphics[width=0.07\textwidth]{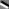}~\includegraphics[width=0.07\textwidth]{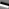}\vspace{4pt}\\\includegraphics[width=0.07\textwidth]{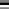}~\includegraphics[width=0.07\textwidth]{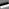}~\includegraphics[width=0.07\textwidth]{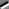}~\includegraphics[width=0.07\textwidth]{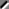}~\includegraphics[width=0.07\textwidth]{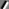}~\includegraphics[width=0.07\textwidth]{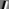}
    \caption{ First derivative of Gaussian (FDOG) Filters of size $9\times 9$ at the orientation difference of $15^{\circ}$.}
    \label{fig:FDOGFilters}
\end{figure}

Despite multiple advantages of the MF-FDOG~\cite{Zhang2010}, the filtering process suffers with inherent limitations. The method considers the parameter $L$ dependent on $\sigma$ due to which filter kernel do not consider neighborhood pixels of vessels into filtering process. To overcome the same, here, we consider that $L$ is independent of $\sigma$ resulting in the consideration of the neighborhood pixels of vessels in the filtering process (see Fig.~\ref{fig:matchFilters} and Fig.~\ref{fig:FDOGFilters}). To make $L$ independent of $\sigma$, the limiting value of $x$ in Equations~(\ref{Eq_MF}) and~(\ref{Eq_FDOG}) are modified to 
\begin{equation}
    |x|\leq L/2
\end{equation}
while keeping all the other parameters same. The designed filters are rotated at different angles to detect the vessels of different orientations. The filters bank of MF and FDOG for $L=9$ and $\sigma =1$ are shown in Fig.~\ref{fig:matchFilters} and Fig.~\ref{fig:FDOGFilters} respectively.

\section{Segmentation}
\label{extSec}
\begin{figure}[!hb]
    \centering
    \includegraphics[width=0.36\textwidth]{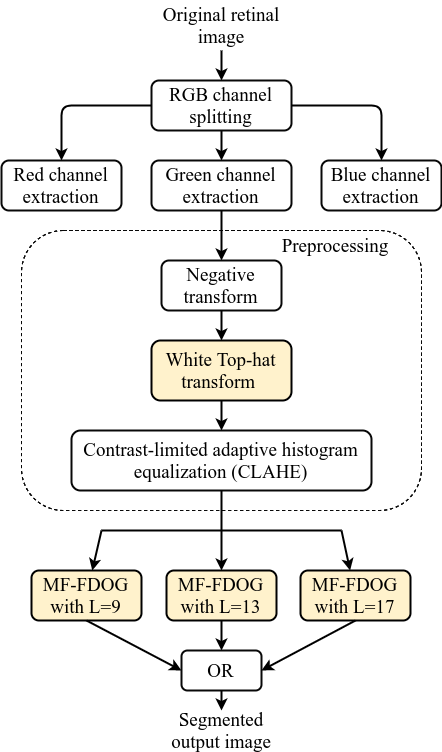}
    \caption{Proposed framework to detect blood vessel}
    \label{fig_framework}
\end{figure}

To extract the blood vessels from retinal fundus images, the proposed framework  is shown in Fig.~\ref{fig_framework}. The main parts of our approach are discussed in following sections.

\subsection{Preprocessing}
Initially, an original RGB color retinal image is separated into red, green, and blue channels where green channel image have higher contrast compare to other two channel images. In green channel image, the blood vessel pixels are visible and easily distinguishable from the background pixels. Because in our eyes, lens pigments absorb light colors differently~\cite{Zhao2014}. Therefore, red vessels in color retinal image are more visible in the green channel image. The green channel image is inverted followed by superimposed with fundus mask  to make the blood vessel brighter and keep the focus on the region of interest . In image filtering, boundary pixels have a grate importance which may deteriorate the segmentation performance if they are not taken care before filtering process. Therefore, to deal with boundary pixels, inverted green channel image are fake padded radially using boundary pixels (shown in Fig.~\ref{FakePad}). Thereafter, white top-hat transformation is used to enhance only the blood vessels. Usually, vessels have small thickness compared to the other structures in retinal images such as exudates, fovea, optic disc, etc. Therefore, a disc shaped structuring element having diameter at least equal to the diameter of the thickest blood vessel in retinal image is preferred for top-hat transformation. The top-hat transformation highlight the object smaller than structuring element and suppress the other structures.
The white top-hat transformed image can be obtained as
\begin{equation}
{T_w}({I_g}) = {I_g} - {I_g} \circ s
\end{equation}
where, $I_g$ is inverted green channel image, $s$ is the structuring element, and $\circ$ denotes the morphological opening operator.  
\begin{figure}[h]
\centering
\includegraphics[width=0.14\textwidth]{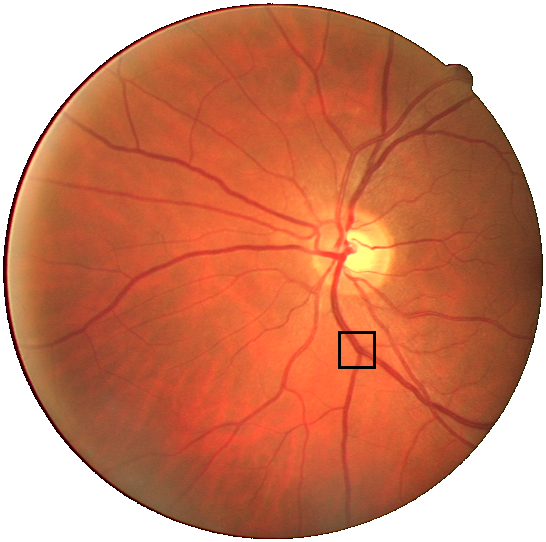}~~
\includegraphics[width=0.14\textwidth]{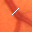}~~
\includegraphics[width=0.14\textwidth]{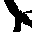}
\caption{Width of widest blood vessel: (a) Marked region on retinal image, (b) Zoomed
vessel of marked region (vessel width around 7-11 pixels), and (c) Ground truth of
zoomed vessel.}
\label{fig_vesselwidth}
\end{figure}
In this paper, the diameter of the structuring element is
chosen 11 pixels wide as the maximum width of the blood vessel is less than 11
pixels. For the DRIVE database, the width of the blood vessel varies in the range
of 1-10 pixels~\cite{Fathi2013}. Fig.~\ref{fig_vesselwidth} shows the width of a wide blood vessel in the retinal
image as 7 pixels wide. The diameter of the structuring element may vary for
the different database having different image resolution. After applying top-hat transformation, blood vessel have poor contrast compare to background. Therefore, contrast limited adaptive histogram equalization (CLAHE) is applied for further image enhancement.
\subsection{Blood vessel extraction}
\begin{figure}[!h]
\centering
\subfigure[]{\includegraphics[height=2.6cm]{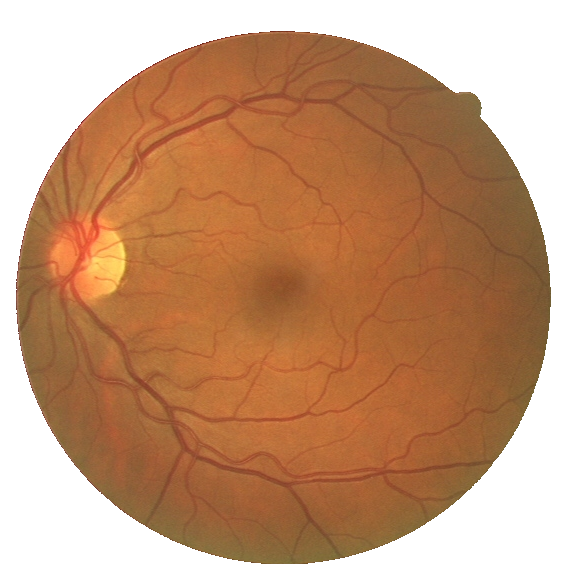}}~~
\subfigure[]{\includegraphics[height=2.6cm]{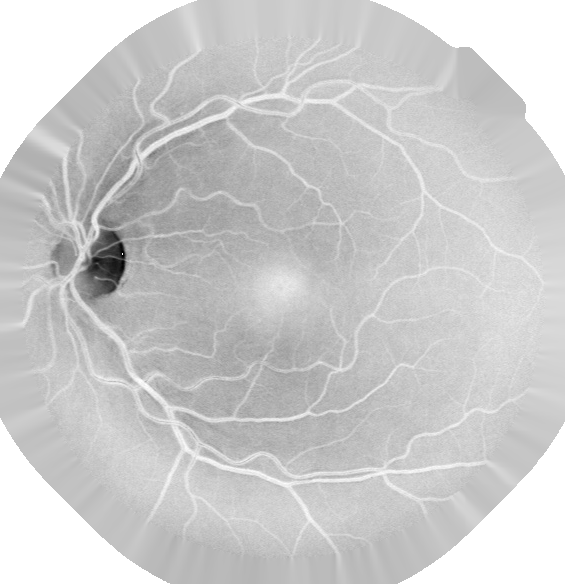}\label{FakePad}}~~
\subfigure[]{\includegraphics[height=2.6cm]{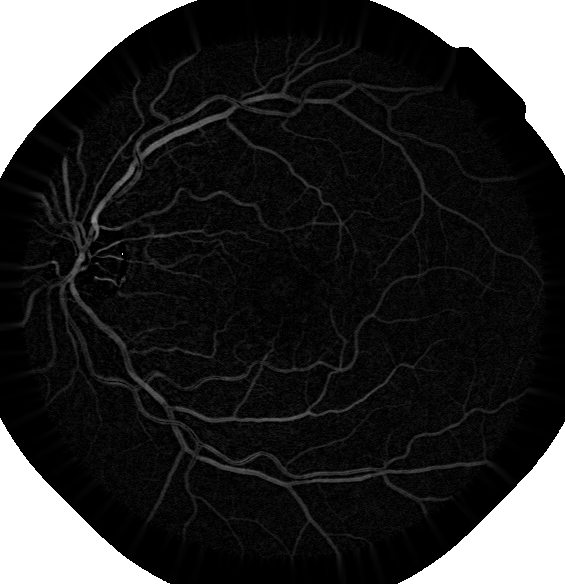}\label{imgGTopHat}}\\
\subfigure[]{\includegraphics[height=2.6cm]{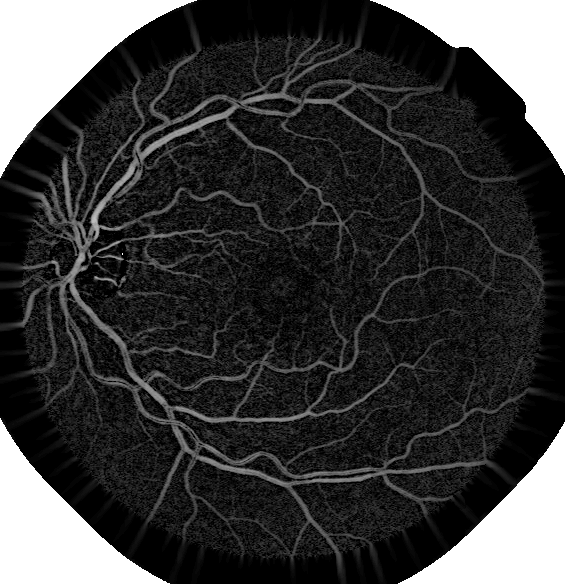}\label{imgClahe}}~~
\subfigure[]{\includegraphics[height=2.6cm]{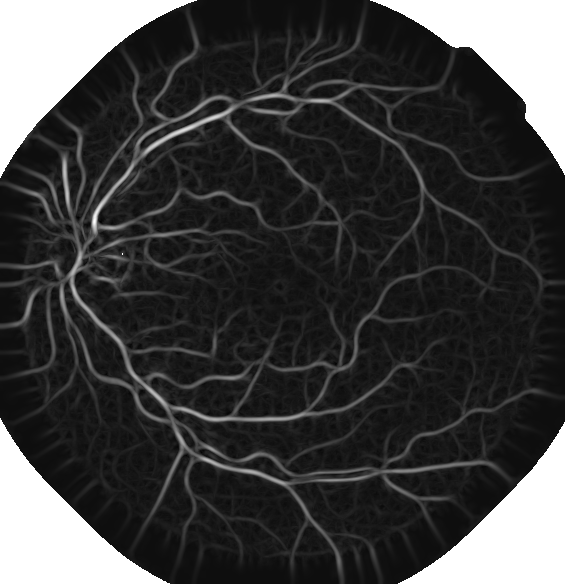}\label{binary}}~~
\subfigure[]{\includegraphics[height=2.6cm]{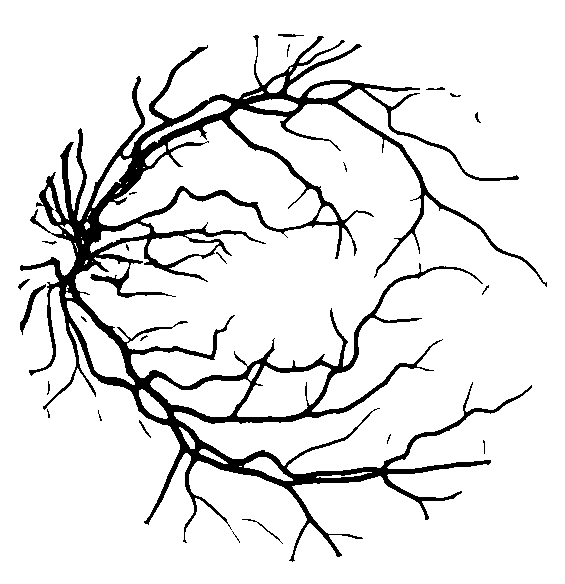}\label{outpurRes}}
\caption{Processing results at each intermediate steps of the proposed algorithm. (a) Original test image, (b) Inverted green channel image with padding, (c) white top-hat transformed image of inverted green channel image, (d) CLAHE processed image, (e) Modified MF response, and (f) Binary blood vessel map over the mask.}
\label{Fig_completStep}
\end{figure}

After preprocessing, enhanced images is further processed through bank of filters obtained using the proposed modified multiscale MF-FDOG. In image processing, filtering operation on edge pixels always degrade the performance of many algorithm. To handle the boundary pixels near field of view (FOV) edges, the retinal image is radially padded using boundary pixels of retina inside FOV in radial direction. The modified MF-FDOG method discussed in section~\ref{ModMFFDOG} is applied on radially padded retinal image (Fig.~\ref{FakePad}). Matched filters of different scales are used to detect thick along with thin vessels. All intermediate steps of the proposed framework (with the help of an example image from DRIVE database) are shown in Fig.~\ref{Fig_completStep}.
\begin{figure*}[!htbp]
    \centering
    \includegraphics[height = 4cm]{img_01_test.png}~~
    \includegraphics[height = 4cm]{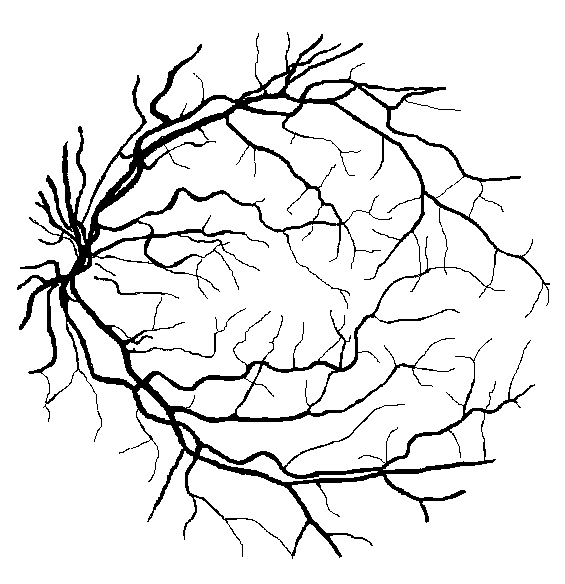}~~
    \includegraphics[height = 4cm]{out_01_test.png}\\
    \includegraphics[height = 4cm]{img_02_test.png}~~
    \includegraphics[height = 4cm]{groundTruth_02_test.png}~~
    \includegraphics[height = 4cm]{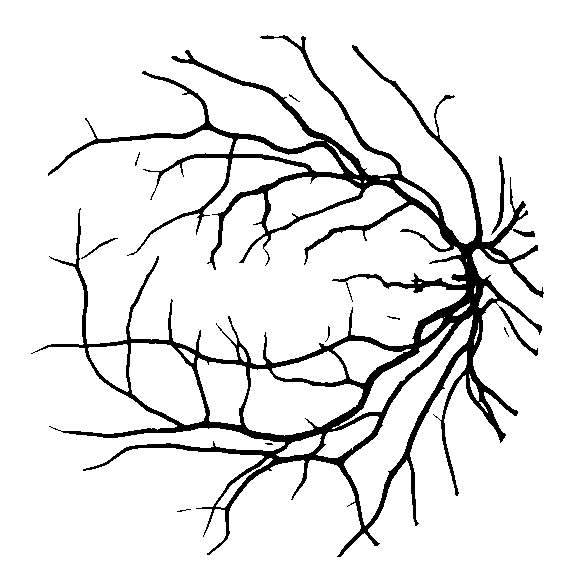}\\
    \includegraphics[height = 4cm]{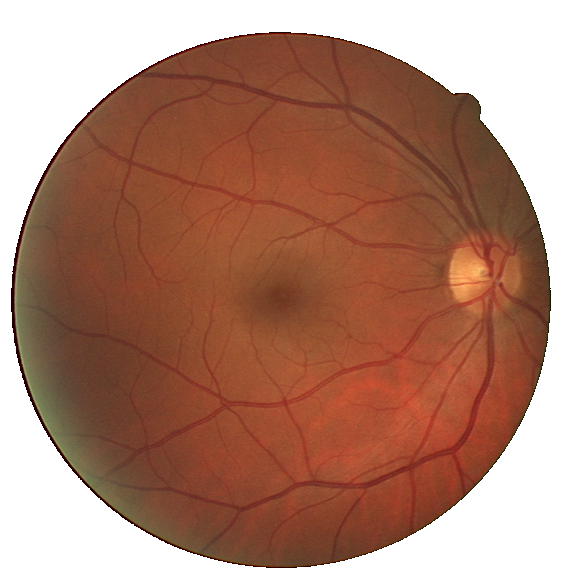}~~
    \includegraphics[height = 4cm]{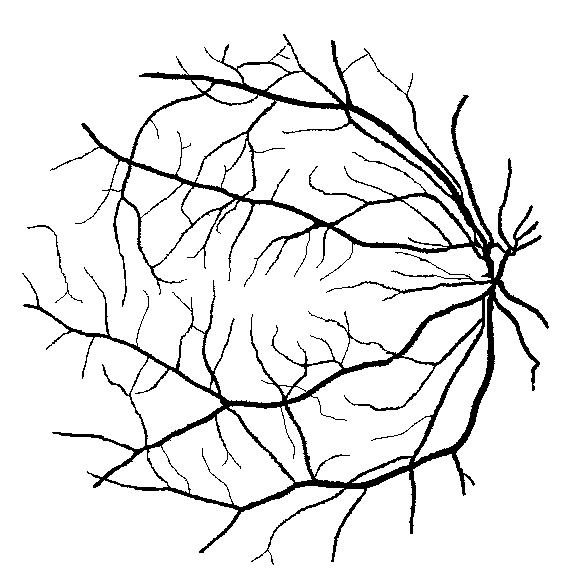}~~
    \includegraphics[height = 4cm]{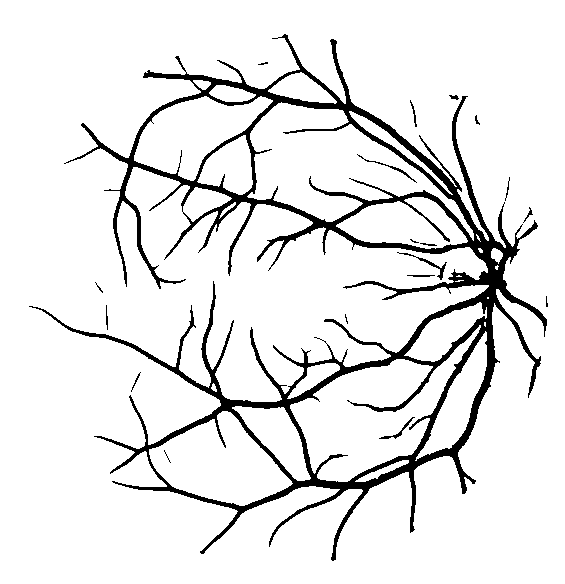}
    \caption{Segmentation result of retinal images from the DRIVE  database. First column: original image, Second column: golden standard ground truth, Third column: segmented vessel using the proposed framework.}
    \label{Fig_SegOutput}
\end{figure*}

\section{Results and Discussion}
\label{resultsec}
\subsection{Database}
In this study, the retinal fundus images are obtained from DRIVE database~\cite{Staal2004}. The database is constructed under diabetic retinopathy screening program conducted in Netherlands. For the database, 40 images have been randomly selected from the total of 400 diabetic patients that have been screened. Each image was captured using a Canon CR5 non-mydriatic 3CCD camera with a 45 degree FOV having 8 bit per color plane of resolution $768\times 584$ pixels. The
results of the manual segmentation (also called gold standard ground truth) are available for the database.

\subsection{Results}
Out of the selected 40 images, 20 images are kept for training and 20 for testing. The proposed framework is evaluated on only test images because the framework is fully unsupervised. Three different scales, $L = $9, 13, and 17 along with $\sigma =$ 1, 1.5, and 2 are used respectively for multiscale MF-FDOG filter design. A bank of MF-FDOG filter kernels are obtained by rotating the kernels at $15^\circ$ increment to obtain 12 different kernels at different orientations. The retinal images from the test set are filtered using those kernels to identify the vessel at different orientation. The segmented blood vessels obtained using the proposed framework are provided along with its gold standard ground truth in Figure~\ref{Fig_SegOutput}. The corresponding results are evaluated and compared with competitive algorithms using three different metrics: accuracy ($Acc$), sensitivity ($Se$), specificity ($Sp$) (see table~\ref{AllResults}). Accuracy, sensitivity, and specificity are calculated using false positive ($X$), false negative ($Y$), true positive ($Z$), and true negative ($W$) values. $Z$ denotes the number of pixels correctly identified as vessel pixels, and $X$ denotes the number of pixels belongs to the background but wrongly identified as vessel pixel. $W$ represents the number of pixels correctly identified as background pixels, and $Y$ represents the number of pixels belongs to the vessels but incorrectly assigned to background pixels. The evaluation metrics are measured using the following mathematical expressions

\begin{equation}
Acc =\frac{Z+W}{Z+X+Y+W}, ~Se = \frac{Z}{Z+Y},~Sp = \frac{W}{X+W}.
\nonumber
\end{equation}

The performance metrics for 20 test images of DRIVE database are tabulated in Table~\ref{AllResults}. The performance metrics illustrate that average segmentation accuracy, sensitivity, and specificity of the presented technique are 94.34\%, 71.93\%, and 97.64\% respectively. Our method resulted in a highly precise identification (deviation in accuracy, $\sigma=0.0049$). Further, Fig.~\ref{Fig_SegOutput} shows that our proposed algorithm is competent in identifying thick as well as thin  vessels in the presence of exudates, fovea, and optic disc. In addition, we provide a comparison table (Table~\ref{compMethod}) of the presented algorithm with different competing methods. Indeed, the results show that our proposed algorithm outperforms many unsupervised methods on DRIVE database concerning accuracy and specificity.

\begin{table}[!h]
\caption{Segmented outcome of the presented work on the DRIVE test database}
\begin{center}
\begin{tabular}{|C{1cm}|C{1.5cm}|C{1.5cm}|C{1.5cm}|}
\hline
\multicolumn{1}{|l}{Image} & Se & Sp & Acc \\ \hline
 1     & 0.8015 & 0.9656 & 0.9441 \\\hline
    2     & 0.7598 & 0.9801 & 0.9471 \\\hline
    3     & 0.7281 & 0.9702 & 0.935 \\\hline
    4     & 0.6206 & 0.9909 & 0.9415 \\\hline
    5     & 0.6611 & 0.9873 & 0.9431 \\\hline
    6     & 0.6729 & 0.9814 & 0.9379 \\\hline
    7     & 0.6779 & 0.9784 & 0.9386 \\\hline
    8     & 0.6837 & 0.9749 & 0.9383 \\\hline
    9     & 0.6959 & 0.9807 & 0.9472 \\\hline
    10    & 0.6759 & 0.9844 & 0.9475 \\\hline
    11    & 0.6985 & 0.9748 & 0.939 \\\hline
    12    & 0.7612 & 0.9681 & 0.9422 \\\hline
    13    & 0.6654 & 0.9832 & 0.9382 \\\hline
    14    & 0.7784 & 0.9675 & 0.9452 \\\hline
    15    & 0.7439 & 0.9748 & 0.9508 \\\hline
    16    & 0.6908 & 0.9815 & 0.9435 \\\hline
    17    & 0.711 & 0.9736 & 0.9412 \\\hline
    18    & 0.7591 & 0.9683 & 0.9443 \\\hline
    19    & 0.8306 & 0.9739 & 0.9567 \\\hline
    20    & 0.7688 & 0.9689 & 0.9475 \\\hline
\multicolumn{1}{|l|}{Average} & 0.7193 & 0.9764 & 0.9434 \\\hline
\multicolumn{1}{|l|}{Std. deviation} & 0.0515 & 0.0068 & 0.0049 \\ \hline
\multicolumn{4}{L{7.5cm}}{Se - sensitivity, Sp - specificity, Acc - accuracy}
\end{tabular}
\end{center}
\label{AllResults}
\end{table}

\begin{table}[!h]
\caption{Comparative analysis of the presented work with existing approach on the DRIVE test data set with respect to golden standard ground truth image}
\begin{center}
\begin{tabular}{|L{2.8cm}|C{1.3cm}|C{1.3cm}|C{1.3cm}|}
\hline
\multicolumn{4}{|L{8cm}|}{\textbf{Unsupervised Methods}}\\\hline
\multicolumn{1}{|l}{Methods}     & Se    & Sp   & Acc  \\ \hline\hline

2nd observer                    & 0.7760 & 0.9725 & 0.9473  \\
Chaudhuri \textit{et al.}\cite{Chaudhuri1989}$^{*}$ &    0.6168    &    0.9741    & 0.8773   \\ 
Zhang \textit{et al.}~\cite{Zhang2010} &   0.7120   &   0.9724   & 0.9382         \\
 Gou \textit{et al.}
\cite{Gou2018} &       0.7526 &    0.9669    & 0.9393          \\ 
\textbf{Presented method}                 & 0.7193 & 0.9764 & 0.9434      
\\ 
\hline
\multicolumn{4}{L{8cm}}{$^{*}$results are taken from ~\cite{Niemeijer2004}}\\
\end{tabular}
\end{center}
\label{compMethod}
\end{table}

\section{Conclusion}
\label{conclusion}
To summarize, here, we have performed the blood vessel extraction on a set of 20 test images of DRIVE database. We found that the top-hat transform followed by CLAHE is a comparatively effective approach to eliminate the structure other than blood vessels in the preprocessing stage. Further, the modified match filter parameters improved the MF-FDOG filters response. Indeed, the proposed framework provides a better average accuracy (94.34\%) and specificity (97.64\%). Moreover, the method can be extended to precisely estimate the object having Gaussian distribution.
\bibliography{icamlBib.bib} 
\bibliographystyle{ieeetr}
\end{document}